# Identification of parameters in amplitude equations describing coupled wakes.

José María Fullana[1], Patrice Le Gal[2], Maurice Rossi[1]
and Stéphane Zaleski[1]

1. *Laboratoire de Modélisation en Mécanique, CNRS URA 229,*
*Université Pierre et Marie Curie, 4 place Jussieu*
*75005 Paris, France. e-mail: zaleski@lmm.jussieu.fr*

2. *Institut de Recherche sur les Phénomènes Hors Équilibre,*
*12 avenue du Général Leclerc,*
*Université d'Aix-Marseille II, CNRS–UMR 138*
*Marseille, France*

We study the flow behind an array of equally spaced parallel cylinders. A system of Stuart-Landau equations with complex parameters is used to model the oscillating wakes. Our purpose is to identify the 6 scalar parameters which most accurately reproduce the experimental data of Chauve and Le Gal [Physica D **58**, pp 407–413, (1992)]. To do so, we perform a computational search for the minimum of a distance $\mathcal{J}$. We define $\mathcal{J}$ as the sum-square difference of the data and amplitudes reconstructed using coupled equations. The search algorithm is made more efficient through the use of a partially analytical expression for the gradient $\nabla \mathcal{J}$. Indeed $\nabla \mathcal{J}$ can be obtained by the integration of a dynamical system propagating backwards in time (a backpropagation equation for the Lagrange multipliers). Using the parameters computed via the backpropagation method, the coupled Stuart-Landau equations accurately predicted the experimental data from Chauve and Le Gal over a correlation time of the system. Our method turns out to be quite robust as evidenced by using noisy synthetic data obtained from integrations of the coupled Stuart-Landau equations. However, a difficulty remains with experimental data: in that case the several sets of identified parameters are shown to yield equivalent predictions. This is due to a strong discretization or "round-off" error arising from the digitalization of the video images in the experiment. This ambiguity in parameter identification has been reproduced with synthetic data subjected to the same kind of discretization.

# 1 Introduction

When analyzing chaotic or turbulent dynamical systems one is often faced with the problem of choosing optimal models. For instance, there are several models of turbulence, such as the Reynolds stress, eddy viscosity or $k$–$\epsilon$ models, that are frequently used by practitioners. One would obviously like to make an optimal choice among this wealth of models. There are two aspects to this: to find the general functional form of the model, and to find the parameters for a given form. In this paper we report our experience with such a *parameter identification problem*.

The experimental system consists in a grid of 16 evenly spaced cylinders placed transversally to a laminar incoming flow. For a single cylinder, when the Reynolds number $Re = \frac{Ud}{\nu}$ — based on the upstream velocity $U$, cylinder diameter $d$ and viscosity $\nu$— exceeds a critical value $Re_c \sim 46$, the celebrated Bénard-Von Karman vortex street sets in. In this regime the wake of the cylinder behaves like a nonlinear oscillator. When an array of cylinders is used as in ref. [1, 2, 3] one obtains instead a chain of oscillators. Such systems have been studied theoretically for some time now (for some recent references see [4]). There are many examples in which they display irregular oscillations just as seen in Figure 1a. The experiment of ref. [1, 2] thus provides a particularly simple hydrodynamical example of spatiotemporal chaos.

The Hopf bifurcation of a single cylinder may be described [5] by a Stuart-Landau equation with complex parameters

$$\partial_t A = \sigma A - l|A|^2 A \tag{1}$$

where $\sigma = r + i\omega$, $l = l_r + il_i$, and $A$ is a complex amplitude related to the observations of the cross-stream deviations of the wakes through its real part. This equation is a normal form for the Hopf bifurcation. Now in the experimental system, the individual wakes with amplitude $A_i$ will undergo some degree of coupling with their neighbors. The simplest possible form for this coupling is a linear finite-difference operator [6]

$$\tau_0 \partial_t A_i = \sigma A_i + g(A_{i+1} + A_{i-1} - 2A_i) - l|A_i|^2 A_i \tag{2}$$

where $g = g_r + ig_i$ is a new complex parameter. Boundary and initial conditions will be discussed later.

Although the three complex parameters $\sigma, l$ and $g$ may in principle be obtained from an analysis of the Navier-Stokes equations, in practice this is a formidable problem, requiring extensive numerical simulations even for a single cylinder. On the other hand, simply searching for the parameters that offer the best possible match to the data requires only small-scale computations with the methods we describe below. To give a more precise definition of our problem, we start with a set of 16 times series $V_{ik} = V_i(t_k)$, where the index $i$ indicates the cylinder and the $t_k$ are the measurement times. We connect the model solutions $A_i(t_k)$ and the real data $V_{ik}$ with the assumption

$$V_{ik} = \text{Re}(A_i(t_k)) + W_{ik} \tag{3}$$

where $W_{ik}$ is a decorrelated noise. Equation (3) expresses the ideal situation where the measurement process adds a Gaussian, independent error to each data value. In what follows however, we will discuss more complex measurement processes (see equation (29) below). Because of these doubts about the validity of equality (3) it is useful to test our parameter identification procedure on a *synthetic* set of data generated directly from a numerical integration of equations (2) and (3). This use of synthetic data offers striking evidence of the role of experimental artifacts such as large noise amplitudes, or other peculiarities of the measurement process.

The selection of an appropriate parameter identification method is a still largely open question. While parameter identification, or more generally the tackling of inverse problems, is classical in various fields of science such as robotics or geophysical research, there is only one attempt we know of to deal with a spatially extended, chaotic dynamical system such as equation (2). Chauve and Le Gal [2] used the proper orthogonal decomposition (POD) to analyze data from the cylinder wake experiment. They were able to extract all the parameters except $g_r$ using the POD. With these parameter values and *ad hoc* values of $g_r$ they produced simulations that were roughly similar to the experimental data, but with incorrect phase relations. While the experimental data show that neighboring wakes are oscillating in opposition to each other, the POD yields wakes oscillating in phase with each other.

In this paper we suggest another identification technique, which is based on an optimization procedure. The optimization consists of a search in the space of all solutions of equation (2) obtained when the parameters are varied. We define a distance $\mathcal{J}$ between the experimental data and a given solution. We then attempt to locate values of the parameters yielding the minimum $\mathcal{J}$. This search is usually difficult, because the investigated space is large and local minima abound. Among the many methods that have been proposed, one of the simplest is to perform a descent in the direction of the gradient which is estimated numerically. (Much more sophisticated methods may be found such as simulated annealing [7] or multigrid methods [8, 9]). We chose here to employ a more sophisticated descent method (the PLMA method described in the Appendix). The main difference with the naive approach lies in the fact that the gradient is not estimated numerically. Instead it is obtained using a technique of backpropagation resulting in an adjoint dynamical system which propagates backwards in time. Although the the field of chaotic model reconstruction has been very active [10, 11, 12], the backpropagation technique is relatively new. We note that it has been used in ref. [13] for *orbit* reconstruction, but not for model reconstruction or, as we do here, for parameter estimation .

In Section 2, we present the experimental setup and data. Section 3 introduces the theoretical model (basically equation (2)) that we take throughout this article as a working hypothesis. In Section 4 we describe the method used for parameter identification, including the definition of the cost function or distance $\mathcal{J}$ and we describe the backpropagation equations for the Lagrange multipliers. Finally

Section 5 contains our numerical results for real and synthetic data sets.

## 2 Experimental Data

Although the data set we analyze in this paper is different from the one used in [1, 2], there are no differences in experimental setup. The 16 cylinders are located in a hydrodynamic water channel. Each cylinder has a 2 mm diameter and is 200 mm long. The cylinder axes are 8 mm apart. Data were gathered above the onset of oscillations at $Re = 80$. Figure 1a gives a visualization of the 16 coupled wakes obtained by the emission of a dye streak at the back of each cylinder. Video images of the wakes are then digitized on 512×512 pixels. The line of pixels located 12 mm downstream of the cylinder's axes is recorded every $T = 40$ ms, yielding data similar to Fig. 1b. These digitized lines are then processed to find the location of the dye streaks. In this first round of data processing, one wants the signal to pass near the maxima of intensity on the videoline. Pixels may have 256 intensity values. Thus there may be several relative maxima or adjacent pixels with the same maximal intensity. The value of the signal is then selected by a procedure that maximizes the smoothness of the signal. Details are given in [14]. The resulting deviation from the centerline position defines the data $V_i(t_k)$ plotted on figure 2a (with $t_k = kT$). Some of the data used in this article will be available at the ftp site `ftp.jussieu.fr` in the directory `jussieu/labos/lmm/Wakes`.

## 3 The Theoretical Model

It is convenient to rewrite equation (2) in the form

$$\frac{d\mathbf{A}}{dt} = \mathbf{H}(\mathbf{A}; \mathbf{a}) \qquad (4)$$

where $\mathbf{A}(t) = (A_i(t))_{i=1,N}$ is a vector of $N$ complex amplitudes, the parameter vector is noted $\mathbf{a} = (r, \omega, l_r, l_i, g_r, g_i)$, and $\mathbf{H}$ is the vector field defined by

$$H_i(\mathbf{A}; \mathbf{a}) = (r + i\omega)A_i + g(A_{i+1} + A_{i-1} - 2A_i) - l|A_i|^2 A_i \qquad (5)$$

Note that cylinders 1 and $N$ pose a special problem since they lie at the ends of the chain. A simplifying hypothesis consists in defining 2 "virtual cylinders" outside the array of the $N = 16$ real cylinders for which a zero amplitude is imposed. Thus

$$A_0 = A_{N+1} = 0. \qquad (6)$$

The above equation plays the role of a boundary condition for our problem.

To analyze the behavior of the system, let us however consider space-periodic solutions, neglecting for a while the effect of the boundary conditions. For $r - 4g_r > 0$ and $l_r > 0$, equation (4) has plane wave solutions of wave number $q$

$$A_i = B e^{\mathrm{i}(\Omega t + qi)} \tag{7}$$

where

$$B = \left( \frac{r + g_r(2\cos q - 2)}{l_r} \right)^{1/2} \tag{8}$$

and

$$\Omega = \omega + g_i(2\cos q - 2) - l_i B^2. \tag{9}$$

The wavenumber $q$ should obey $r + g_r(2\cos q - 2) > 0$. From these solutions it is easy to grasp the physical meaning of the various parameters. The distance to the instability threshold for in-phase ($q = 0$) oscillations is proportional to $r$, and the amplitude of the oscillations grows with it. The coupling between neighboring oscillators is a discretized version of the differential operator $\nabla^2 A$ that appears in the continuous-space Ginzburg-Landau equation. With positive $g_r$ this term tends to damp plane wave oscillations with a non uniform phase ($q \neq 0 \mod 2\pi$). With $g_r < 0$ oscillations with a phase shift between cylinders are amplified, and the maximum amplification rate is achieved when $q = \pi$. The parameters $\omega$, $g_i$ and $l_i$ control the frequency of the oscillations.

To perform computations, we need to discretize the continuous-time model (4). We hence divide the time window $(t_0, t_0 + j_{\max}T)$ into $M - 1$ time steps $\tau$. Experience shows that $\tau$ must be much smaller than $T$, on the order of $T/100$, to reach accurate results. The discrete system is a simple forward Euler discretization

$$\mathbf{A}_{k+1} = \mathbf{A}_k + \tau \mathbf{H}(\mathbf{A}_k; \mathbf{a}) \quad \text{for} \quad k = 0, \cdots, M - 1. \tag{10}$$

where $\mathbf{A}_k = \mathbf{A}(t_0 + k\tau)$.

Taking another point of view it is possible to chose a large time step $\tau = T$ and to attempt to minimize the distance $\mathcal{J}$ with the resulting equation although it is an obviously poor approximation to equation (4). In this new point of view it is the discrete model rather than the continuous model which is postulated. This empirical approach is convenient because the discrete times correspond one to one to the data points, and that the computational work needed to integrate the model is scaled down by about 2 orders of magnitude[1].

---

[1] It also allows for a simple type of least-squares fit of the parameters on which we will report elsewhere

# 4 Method

To identify the parameters, we first need to define a distance $\mathcal{J}$ between experimental observations and model predictions. Let $\mathbf{\Phi}(\mathbf{A}_0, t; \mathbf{a})$ be the flow integrated from equation (4) with parameters $\mathbf{a}$, defined so that if $\mathbf{A}_0 = \mathbf{A}(t_0)$, then $\mathbf{A}(t) = \mathbf{\Phi}(\mathbf{A}_0, t - t_0; \mathbf{a})$ is the solution at time $t$. As in equation (3) the observations are related to the real part of the amplitudes $\mathbf{A}(t)$. We let $t_j^{\text{obs}} = jT$ be the times at which observations are made, with $T$ the sampling period of the experiment. We thus define the distance by

$$\mathcal{J}(\mathbf{A}_0, \mathbf{a}; t_0, j_{\max}) = \sum_{j=1}^{j_{\max}} ||\mathbf{V}(t_j^{\text{obs}}) - Re[\mathbf{\Phi}(\mathbf{A}_0, t_j^{\text{obs}} - t_0; \mathbf{a})]||^2 \qquad (11)$$

where $\mathbf{V}(t) = (V_i(t))_{i=1}^{i=N}$ is the $N$-vector of time series of observations, $j_{\max}T$ is the length of the series used for the definition of $\mathcal{J}$, and $||.||^2$ is the distance squared defined by $||\mathbf{X}||^2 = \sum_i X_i^2$. In expression (11), the time $t_0$ need not be the beginning of the full data set represented on Fig. 2a. Likewise, there is some leeway in the choice of the length $j_{\max}T$. Thus, we may define a distance for each time window $(t_0, t_0 + j_{\max}T)$. For each of these "training windows" the parameter identification is cast as an optimization problem, in which one searches for a global minimum of $\mathcal{J}$ with respect to a set of parameters. In the simplest version of this problem we express it as

$$Find\ \mathbf{a}_{op} \in R^6\ such\ that\ \mathcal{J}(\mathbf{A}_0, \mathbf{a}_{op}) \leq \mathcal{J}(\mathbf{A}_0, \mathbf{a})\ \ \forall \mathbf{a} \in R^6$$

In more complex versions, we may add other parameters that we wish to vary, such as the initial conditions $\mathbf{A}_0$.

In order to solve this problem computationally, we use the PLMA algorithm proposed by Gill and Murray [15] and described in the Appendix. This method provides a minimum of $\mathcal{J}$ when its gradient $\nabla_{\mathbf{a}}\mathcal{J}$ is available. While it is possible to find a gradient direction from several numerical estimates of $\mathcal{J}$, an efficient and accurate computation of the gradient is crucial since identification problems are generally ill-conditioned and lead to a large number of evaluations of the gradient. We thus construct an explicit analytical expression for the gradient $\nabla_{\mathbf{a}}\mathcal{J}$. This expression is derived from the problem defined in *discretized time* as in equation (10). This is a requirement which has been observed in practice to condition the success of the optimization algorithm. The distance thus reads

$$\mathcal{J}_1(\mathbf{a}, \mathbf{A}_0, \mathbf{A}_1, \cdots, \mathbf{A}_M) = \sum_{k=1}^{M} \epsilon_k ||\mathbf{V}_k - \mathbf{X}_k||^2 \qquad (12)$$

where $\mathbf{V}_k$ stands for the vector of experimental amplitudes at time $k\tau$, we decomposed the amplitudes into real and imaginary parts $\mathbf{A} = \mathbf{X} + i\mathbf{Y}$, and we introduced an index $\epsilon_k$ which is equal to one when experimental data are available — that is every $T = 40ms$ — and zero in the opposite case. Finally we have used the notation $\mathcal{J}_1$ — instead of $\mathcal{J}$ — to express the fact that $\mathcal{J}_1$ is a function of the whole time series of amplitudes $\mathbf{A}_0, \mathbf{A}_1, \cdots, \mathbf{A}_M$. When these amplitudes are related by the discrete model (10), we have $\mathcal{J} = \mathcal{J}_1$.

This completely defines the cost function in discrete time, but the dependence of $\mathcal{J}_1$ on $\mathbf{a}$ is implicit through its dependence on the variables $\mathbf{A}_k$. This difficulty can be handled using the Lagrange multipliers method. We hence recall a basic fact about constrained differentiation:

Let $\mathcal{J}_1(a, x)$ be a function of $R^n \times R^m \to R$. Consider the set of $m$ constraints $c(a, x) = 0$ where $c$ is a function of $R^n \times R^m \to R^m$. Assume that the implicit function theorem applies so that these constraints define $x$ as a function $x(a)$ of $a$. Let $\mathcal{J}$ be the function of $R^n \to R$ defined by $\mathcal{J}(a) = \mathcal{J}_1(a, x(a))$. Call Lagrangian the function of $R^n \times R^m \times R^m \to R$ defined by

$$\mathcal{L}(a, x, p) = \mathcal{J}_1(a, x) + c(a, x) \cdot p, \quad (13)$$

where the components of $p \in R^m$ are called Lagrange multipliers. Under these conditions, if

$$\nabla_p \mathcal{L} = 0 \quad (14)$$

and

$$\nabla_x \mathcal{L} = 0 \quad (15)$$

then

$$\nabla_a \mathcal{J} = \nabla_a \mathcal{L}. \quad (16)$$

We use this fact to compute explicitly the gradient of $\mathcal{J}$. We introduce the multipliers $\mathbf{P}_k = (P_{ik})_i$, $\mathbf{Q}_k = (Q_{ik})_i$ for $k = 0, \cdots, M-1$ and $i = 1, \cdots, N$, and the discrete Lagrangian

$$\mathcal{L} = \sum_{k=1}^{M} \left\{ \epsilon_k ||\mathbf{V}_k - \mathbf{X}_k||^2 + [\mathbf{X}_k - \mathbf{X}_{k-1} - \tau \mathbf{F}(\mathbf{A}_{k-1}; \mathbf{a})] \cdot \mathbf{P}_{k-1} \right.$$
$$\left. + [\mathbf{Y}_k - \mathbf{Y}_{k-1} - \tau \mathbf{G}(\mathbf{A}_{k-1}; \mathbf{a})] \cdot \mathbf{Q}_{k-1} \right\}. \quad (17)$$

where we have taken $\mathbf{H} = \mathbf{F} + i\mathbf{G}$. For this Lagrangian, relation (14) is equivalent to the direct equation (10), while relation (15) yields the equations

$$\frac{\partial \mathcal{L}}{\partial X_{ik}} = 0 \quad \frac{\partial \mathcal{L}}{\partial Y_{ik}} = 0, \quad (18)$$

where these equations hold for all amplitudes $A_{ik}$ which are not fixed by boundary or initial conditions, that is for $k = 1, \cdots, M$ and $i = 1, \cdots, N$. After some algebraic manipulation, conditions (18) may be put in the form

$$\begin{aligned} \mathbf{P}_{M-1} &= 2\tau(\mathbf{V}_M - \mathbf{X}_M), \\ \mathbf{Q}_{M-1} &= 0, \end{aligned} \quad (19)$$

(notice $\epsilon_M = 1$) and

$$\begin{aligned} \mathbf{P}_{k-1} &= \mathbf{P}_k + \tau[H_{11}\mathbf{P}_k + H_{12}\mathbf{Q}_k - 2\epsilon_k(\mathbf{V}_k - \mathbf{X}_k)], \\ \mathbf{Q}_{k-1} &= \mathbf{Q}_k + \tau[H_{21}\mathbf{P}_k + H_{22}\mathbf{Q}_k], \end{aligned} \quad (20)$$

for $0 < k < M$, where we introduce the $N \times N$ matrix differential operators $H_{11} = (\partial F_i/\partial X_j)_{ij}$, $H_{12} = (\partial F_i/\partial Y_j)_{ij}$, $H_{21} = (\partial G_i/\partial X_j)_{ij}$, $H_{22} = (\partial G_i/\partial Y_j)_{ij}$ evaluated at the discrete time $k$. The action of these operators may be written explicitly as

$$\begin{aligned} H_{11,ij} P_{j,k} &= \left[r - l_r(3X_{ik}^2 + Y_{ik}^2) + 2l_i X_{ik} Y_{ik}\right] P_{ik} \\ &\quad + g_r(P_{i-1,k} + P_{i+1,k} - 2P_{ik}), \end{aligned} \quad (21)$$

$$\begin{aligned} H_{12,ij} Q_{j,k} &= \left[\omega - l_r 2X_{ik} Y_{ik} - l_i(3X_{ik}^2 + Y_{ik}^2)\right] Q_{ik} \\ &\quad + g_i(Q_{i-1,k} + Q_{i+1,k} - 2Q_{ik}), \end{aligned} \quad (22)$$

$$\begin{aligned} H_{21,ij} P_{j,k} &= \left[-\omega - l_r 2X_{ik} Y_{ik} + l_i(3Y_{ik}^2 + X_{ik}^2)\right] P_{ik} \\ &\quad - g_i(P_{i-1,k} + P_{i+1,k} - 2P_{ik}), \end{aligned} \quad (23)$$

$$\begin{aligned} H_{22,ij} Q_{j,k} &= \left[r - l_r(3X_{ik}^2 + Y_{ik}^2) - 2l_i X_{ik} Y_{ik}\right] Q_{ik} \\ &\quad + g_r(Q_{i-1,k} + Q_{i+1,k} - 2Q_{ik}). \end{aligned} \quad (24)$$

Equations (20) are called the backpropagated equations, since they may be solved iteratively by decrementing the discrete time $k$, using equations (19) as initial conditions. The gradient of the Lagrangian with respect to the parameters, which by (16) is also the gradient of our cost function, reads

$$\frac{\partial \mathcal{L}}{\partial a_i} = -\sum_{k=0}^{M-1} \left[\mathbf{P}_k \cdot \frac{\partial \mathbf{F}(\mathbf{A}_k; \mathbf{a})}{\partial a_i} + \mathbf{Q}_k \cdot \frac{\partial \mathbf{G}(\mathbf{A}_k; \mathbf{a})}{\partial a_i}\right]. \quad (25)$$

This gradient may be computed in the following way: (i) For a given value of the parameters and initial conditions, compute the $\mathbf{A}_k$ from the direct equations (10). (ii) Compute the $\mathbf{P}_k$ and $\mathbf{Q}_k$ from the backpropagated equations. (iii) Compute the gradient using (25).

It is possible to generalize this procedure to the yield the gradient of the distance $\mathcal{J}$ with respect to other quantities. For instance, the imaginary parts of the initial amplitudes may be considered as parameters of the problem. Differentiating the Lagrangian, one obtains

$$\frac{\partial \mathcal{L}}{\partial Y_{i0}} = -\frac{1}{\tau}Q_{i0} - \frac{\partial G_j(\mathbf{A}_0;\mathbf{a})}{\partial Y_{i0}}Q_{j0} - \frac{\partial F_j(\mathbf{A}_0;\mathbf{a})}{\partial Y_{i0}}P_{j0}. \qquad (26)$$

Likewise, it would be possible to consider parameterized boundary conditions to replace (6) although we have not performed this kind of optimization in the work reported here. Similarly, we always use $\mathbf{X}_0 = \mathbf{V}_0$ for the real part of the initial conditions, although the large error in the data acquisition may justify an optimization step for $\mathbf{X}_0$ as well. When the optimization algorithm is not used to pick $\mathbf{Y}_0$, we use a Hilbert transform to recover the imaginary part, as in [5]. Specifically, for a signal $x(t)$ which is assumed to be the real part of an analytical signal $z_x(t)$ one has [16]

$$z_x(t) = x(t) - \mathrm{i}\, TH[x(t)]$$

where

$$TH[x(t)] = x(t) * \frac{1}{\pi t} = V.P. \int_{-\infty}^{\infty} \frac{x(\theta)}{t - \theta} d\theta$$

where V.P. is the Cauchy principal value. This transformation is strictly valid when the time series has infinite length. For a finite-length time series obtained from simulations, our numerical experiments have shown that the Hilbert transform poorly converges to the true imaginary part of the amplitudes. However, if the Hilbert transform is performed on the whole data series of length 256, but the first and last 30 points are excluded, a reasonable approximation to the imaginary parts is obtained. It is worthwhile to stress this last point, since we have found that it is not possible to obtain satisfactory results without this elimination of a string of data at the beginning and the end of the time series.

# 5   Results

We present here some numerical results pertaining both to experimental data and to artificial or "synthetic" data. The optimization procedures for model (4) were performed in various ways. In a first procedure the initial imaginary parts of the amplitudes are obtained by the Hilbert transform, described above, of the real amplitudes. We have varied the length of the training window $(t_0, t_0 + j_{\max}T)$ and found that the best reconstruction was attained near $j_{\max} = 30$. This corresponds approximately to 2,5 periods of the basic solution and is of the order of the correlation time of the system. On Figure 3 we plot the value of the optimal parameters as the time step $\tau$ of the integration is varied. We observe that the parameters have a well-defined limit as $\tau/t \to 0$. These parameters for the first computed window are shown in Table 1. This first window is the $30-60$ window because of limitations arising from our use of the Hilbert transform. The parameter values are made non-dimensional using the sampling time $T = 40$ms. Notice that the parameter $r$ is negative, but that the amplitude $B$ of plane waves in phase opposition ($q = \pi$) in equation (8) is still real. The parameter values obtained in the first ($30-60$) training window are used in a numerical simulation of the model (4), which yields the *reconstructed data*. Figure 5a shows the reconstructed data (dotted line) and the experimental data (solid line). The series is plotted over the length of the $30-60$ window. To quantify the error we define the normalized prediction error $e_k$:

$$e_k = ||\mathbf{X}_k - \mathbf{V}_k||^2 / ||(\mathbf{V}_k - \langle \mathbf{V}_k \rangle)||^2$$

Table 1: Estimated values of the parameters

|  | $30-60$ training window | average for all windows | 8 parameters average |
|---|---|---|---|
| $rT$ | -0.0084 | $-0.0060 \pm 0.003$ | $-0.019 \pm 0.016$ |
| $\omega T$ | 0.528 | $0.535 \pm 0.006$ | $0.50 \pm 0.007$ |
| $g_r T$ | -0.00523 | $-0.0045 \pm 0.001$ | $-0.0041 \pm 0.0006$ |
| $g_i T$ | -0.00259 | $-0.0038 \pm 0.0015$ | $-0.0043 \pm 0.001$ |
| $l_r T$ | 0.01707 | $0.029 \pm 0.01$ | $0.15 \pm 0.16$ |
| $l_i T$ | -0.0986 | $-0.081 \pm 0.02$ | $-0.37 \pm 0.05$ |
| $d_r T$ | — | — | $0.41 \pm 0.4$ |
| $d_i T$ | — | — | $0.46 \pm 0.07$ |

The normalized error $e_k$ may be interpreted as follows: when $e_k = 0$ the prediction equals the data while, when $e_k = 1$ the prediction does not approach the data better than the average of the signal. The normalized errors for the first training window are small, indicating that our fit is very satisfactory. Figure 4a present the entire simulation over the full duration for which we have experimental data. Figure 4b is the corresponding power spectrum. We also checked that the characteristic appearance of oscillatory periods with interspersed low amplitude or "laminar" phases is recovered even at times much longer than the correlation time (see Fig. 12b). On the other hand, after the initial correlation time, the error now grows to $\mathcal{O}(1)$ levels. This growth of the error is unavoidable, as it is related to the sensitive dependence to initial conditions.

Another qualitative aspect is that we recover the same phase relationship as in experimental data: one observes that wakes oscillate in phase opposition ($q = \pi$ in equation (7)). This is in stark contrast with the results of the POD analysis of [2] where neither short time predictions nor phase relationships are recovered.

We also attempted a second, rather different procedure for parameter estimation. Instead of estimating the initial imaginary parts $Y_{i0}$ using the Hilbert transform we included, using expressions (26), these imaginary parts in the set of parameters for which we try to optimize. This yields a new estimate of both **a** and $Y_{i0}$. Figures 6 shows the long-time predictions and the short-time error after optimization of these 22 parameters. The prediction error over the training window $(t_0, t_0 + j_{\max}T)$ is somewhat smaller. On the opposite the long time simulation (Fig. (6)a) differs drastically from the data: several cylinders have much smaller amplitudes than in the experiment. This indicates a likely over-fit situation. For the results reported in what follows, we reverted to the estimation of $Y_{i0}$ using the Hilbert transform.

It is also interesting to ask whether we can distinguish between several functional forms for the equation (4). In a preliminary attempt, we extended our model by adding a higher-order nonlinear term. We changed **H** to

$$H_i(\mathbf{A}; \mathbf{a}) = (r + i\omega)A_i + g(A_{i+1} + A_{i-1} - 2A_i) - l|A_i|^2 A_i - d|A_i|^4 A_i \qquad (27)$$

with a new complex parameter $d = d_r + id_i$. We found that the cost function has many relative minima. This makes the optimisation procedure much more difficult, As a preliminary result, we selected the data in the window $(30, 90)$ in which a global minimum is easy to find. This yields the parameters in column 4 of Table 1 and the predictions shown on Figures 7 and 8 . The short-time error is approximately the same. The long-time results show a qualitatively different behavior, with much longer intermittent episodes. Thus it seems that the simplest, 6 parameter model is better.

A test of the consistency of our reconstruction is obtained by shifting the starting point $t_0$ of the training window. Figure 9 shows the variation of the optimized parameters as a function of $t_0$. There is a great deal of scatter in this result. Since the reconstructed signals are always satisfactory, we interpreted this scatter as a degeneracy in the optimized system. To give a simple example of such a degeneracy, if our data were a space-and-time-periodic signal of the form (7), it could be reconstructed with parameters in a codimension-2 set, constrained only by the 2 relations (8) and (9). In other words, all parameters in a 4-dimensional manifold would fit periodic data. This degeneracy can be lifted only by non-periodic episodes in the data. Such episodes however are relatively sparse in our data set. Moreover there is still the possibility of a lower order degeneracy for more complex solutions than the plane waves (7).

To test our optimization method independently of the experimental data set we produced a set of synthetic data. We attempt to mimic as closely as possible the experimental process. (i) First we simulate the model equation (4) with the parameters of Table 1, column 2. We already know that these parameters yield a behavior similar to that of the data. This yields a series of amplitudes $A_{ik}$, from which we extract the real parts $X_{ik}$. (ii) We then average the signal over the sampling period $T$ to approximate the time-averaging produced by the video recording. (iii) We then "round-off" or "discretize" the output signal, by letting it take only a small number of integer values. This round-off mimics the processing of the digitized images, in which the wake positions are located at an integer number of pixels away from the centerline. The first two steps result in the signal $V_{ik}$ defined for every $t_k^{\text{obs}}$

$$\tilde{V}_{ik}^{(1)} = \frac{1}{L} \sum_{n=1}^{L} X_{i,kL+n} + W_{ik} \qquad (28)$$

where $L = T/\tau$ and $W_{ik}$ is a decorrelated noise. The third step is

$$V_{ik} = E\left[\frac{S}{V_{\text{max}}} \tilde{V}_{ik}^{(1)} + 0.5\right] \qquad (29)$$

where $E(x)$ is the largest integer smaller than $x$ and $V_{\text{max}}$ is an upper bound for the simulated data $V_{ik}$. The number $S$ controls how discretized the final signal is: the resulting data $V_{ik}$ may have integer values in the interval $(-S, S)$. This results in a "deterministic" loss of information, which is quite different from the kind of loss of information that occurs with an added noise source. This should be compared with the discreteness of the original data, which were equivalent to integer values in the range $(-5, 5)$. An example of the resulting data and the original experimental data at the same scale are shown on figure 10.

Since the true values $\overline{a}_i$ of the parameters are known for synthetic data, we represent the estimated parameters $\hat{a}_i$ divided by $\overline{a}_i$. Figure 11 shows the variation of parameters with the starting point $t_0$ of the training window. The noise level was adjusted to mimic the experimental data. We let the ratio of the noise standard deviation to the data standard deviation be $\sigma[W_{ik}]/\sigma[X_{ik}] = 0.085$. Even with such relatively high noise levels, only a fraction of the variability on figure 9 is recovered. The parameters are barely affected by the averaging procedure (28), and we have found that most of the error comes from the Hilbert transform and the threshold process and not from the noise. Clearly we have explained only a part of the variability. It is of course possible that the measurement process introduces a larger noise amplitude $\sigma[W_{ik}]$, but other explanations are also possible. For instance, one may imagine a slow drift of the parameter values during the experiment,

Finally, we have attempted to optimize parameters for the "large-time-step model" obtained by letting $\tau = T$ in equations (10). Figures 12a and 13 show the entire original data set and its simulation where experimental initial conditions are used and imaginary parts are calculated by Hilbert transform. Figure 12b shows a long run of the moduli $|A|$. This plot demonstrates that intermittency persists over long times. Similar long time results (not shown here) were obtained in the continuous case. It is quite interesting to note that the accuracy of the reconstruction in the large-time-step model is of the same order as that of the "converged model" in which $\tau \ll T$.

# 6 Conclusion

We have introduced a method for the optimization of parameters appearing in a system of coupled Stuart-Landau equations. The optimized model appears to reproduce the data in a very satisfactory way, making possible both short term predictions and a longer term reproduction of the qualitative features of the data, such as intermittency. The method thus appears promising for the reconstruction of dynamical systems from noisy data.

On the physics side, it appears that the experimental data may be reproduced quite accurately by a continuous-time discrete-space Ginzburg-Landau model. This is perhaps not surprising given the previous success of qualitative comparisons between such systems and experimental data. However, it is to be noted that the experiments are conducted far from the Hopf bifurcation at $Re_c$, where it may be surprising that a theory designed for the vicinity of the bifurcation may still hold. Moreover, it is interesting to compare our estimated values of the parameters with those found in the literature. We notice that the coupling renormalizes the real part $r$ of the growth rate: while $r$ is negative in our reconstructions, the plane wave solutions with $q = \pi$ still grow. It is also interesting to compare the ratio $l_i/l_r$ to previous numerical and experimental estimates. For arrays of cylinders the POD methods of ref. [2, 3] give $l_i/l_r = -1.4$ albeit with undetermined accuracy. For a single, infinite aspect ratio cylinder near threshold (see [17],[18] and references therein), $l_i/l_r = -2.7 \pm 0.1$ was found but this ratio seems to decrease towards $-1$ as the aspect ratio decreases. We find $l_i/l_r = -2.8$, from column 3 of Table 1 but as for the result of [2, 3] we may have a very large error. It is interesting to go byond the fit of parameters for the coupled Landau equation to investigate other models. Our results in this direction are only preliminary. Higher order nonlinearities as in (27) make optimisation much more difficult. They also produce qualitatively different solutions, with longer intermittent episodes. Recently, P. Legal and his coworkers have observed steady states in which the wakes of some cylinders stop oscillating indefinitely. We obtained this behavior for some parameter sets of the extended model (27).

Looking further to the general applicability of our method, It is interesting to discuss the dual nature, spatial and temporal, of our data and models. In the method we have used, nothing requires in principle our data to be extended spatially. Thus the method could be used just as well for dynamical systems with a small number of degrees of freedom. On the other hand, the spatial extension of our system results in the production of much more data than in a small system. In fact, it is possible in theory to increase the number $N$ of interacting oscillators in such systems while keeping the number of parameters constant. In fact, systems in higher dimensionality should be more interesting from that point of view. One would expect the amount of data to grow exponentially, much faster than the number of parameters as the dimensionality is increased.

One difficulty with the present approach is that the parameters cannot be unambiguously determined as they vary with the training window used in the simulations. The origin of this variation is unclear, although part of it may be recovered in trials performed with synthetic data. Determining the cause of this phenomenom is certainly a worthy topic for future research.

The method we describe in this paper may also be a basis for the more ambitious goal to optimize to find the equations of motion themselves rather than simply parameters of a given equation. We have tried to introduce a modest variation of the model by adding a higher order nonlinear term, but a much more systematic search for better models should be possible. We believe that a more thorough experience with such optimization techniques would open the door to model optimization for a large number of spatially extended systems.

## Acknowledgments

We benefited from stimulating and instructive conversations with Carey Bunks and Guy Chavent.

# Appendix

In this appendix, we shortly describe the PLMA algorithm. Let $a_0$ be a given initial guess for the parameters, where the sought parameters minimize the cost function $\mathcal{J}(a)$. Let $k$ denote the current iteration, starting with $k = 0$. Each iteration requires the gradient vector $G_k$ evaluated at the $k$th estimate of the minimum $a_k$. A vector $d_k$ (know as the direction of search) is computed and the new estimate $a_{k+1}$ is given by $a_k + \alpha_k d_k$ where the step length $\alpha_k$ is chosen to minimize the function $\mathcal{J}(a_0 + \alpha_k\, d_k)$. A quasi-Newton method is used to compute the search direction $d_k$ by updating the inverse of the approximate Hessian ($\tilde{H}_k$) and computing

$$d_{k+1} = -\tilde{H}_{k+1} G_{k+1} \qquad (30)$$

The updating formula for the approximate inverse is given by

$$\tilde{H}_{k+1} = \tilde{H}_k - \frac{1}{y_k^T s_k}(\tilde{H}_k y_k s_k^T + s_k y_k^T \tilde{H}_k) + \frac{1}{y_k^T s_k}(1 + \frac{y_k^T \tilde{H}_k y_k}{y_k^T s_k}) s_k s_k^T \qquad (31)$$

where $y_k = G_{k+1} - G_k$ and $s_k = a_{k+1} - a_k = \alpha_k d_k$

The algorithm uses a two-step method (with $\tilde{H}_0$ equal to the identity matrix) described in detail in [15] in which restarts and pre-conditioning are incorporated. The termination criterion involves a pre-defined tolerance $\epsilon_\tau$. The algorithm is halted if the following three conditions are satisfied

i $\mathcal{J}_{k-1} - \mathcal{J}_k < \epsilon_\tau(1 + |\mathcal{J}_k|)$

ii $||a_{k-1} - a_k|| < \epsilon_\tau^{1/2}(1 + ||a_k||)$

iii $||G_k|| < \epsilon_\tau^{1/3}(1 + |\mathcal{J}_k|)$ or $||G_k|| < \epsilon_A$, where $\epsilon_A$ is the absolute error associated with computing the cost function $\mathcal{J}$.

# Figures

for Fig 1a)  send mail

to fullana@lmm.jussieu.fr

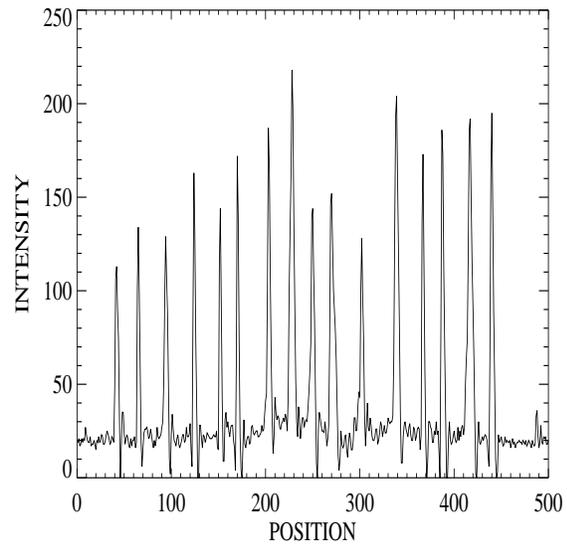

Figure 1: a) Snapshot of the experiment. b) Original data acquisition for fixed time. Peaks locate tracer lines. The deviation of tracer lines from their baseline position defines the experimental wake amplitudes $V_{ik}$.

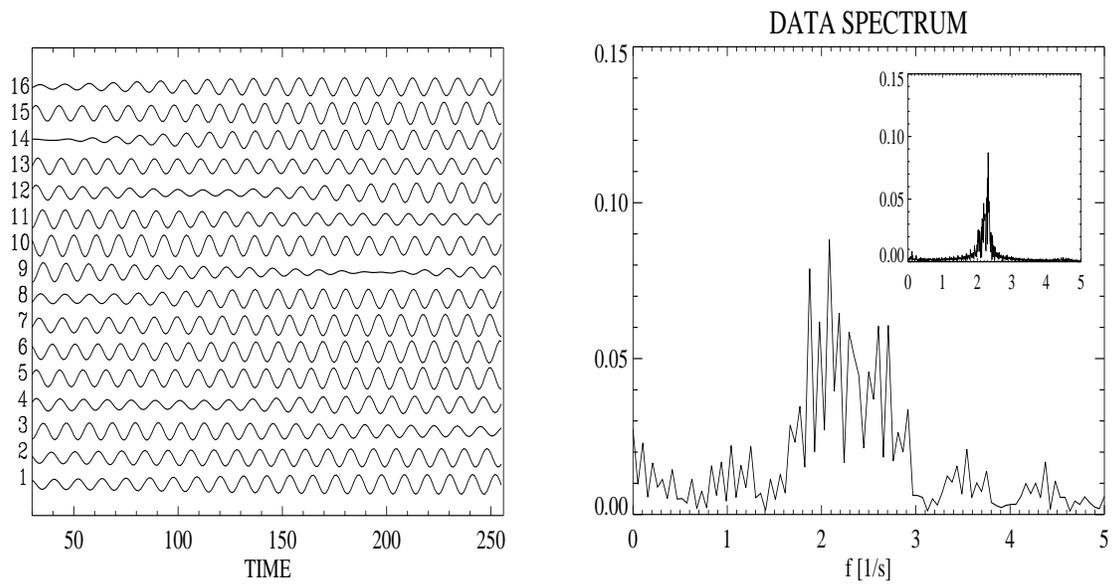

Figure 2: (a) Experimental data composed by 256 original acquisitions. (Showed in the range 30-256) (b) Power spectrum of the original signal computed over the identification window. The inset shows the spectrum computed over the entire time series.

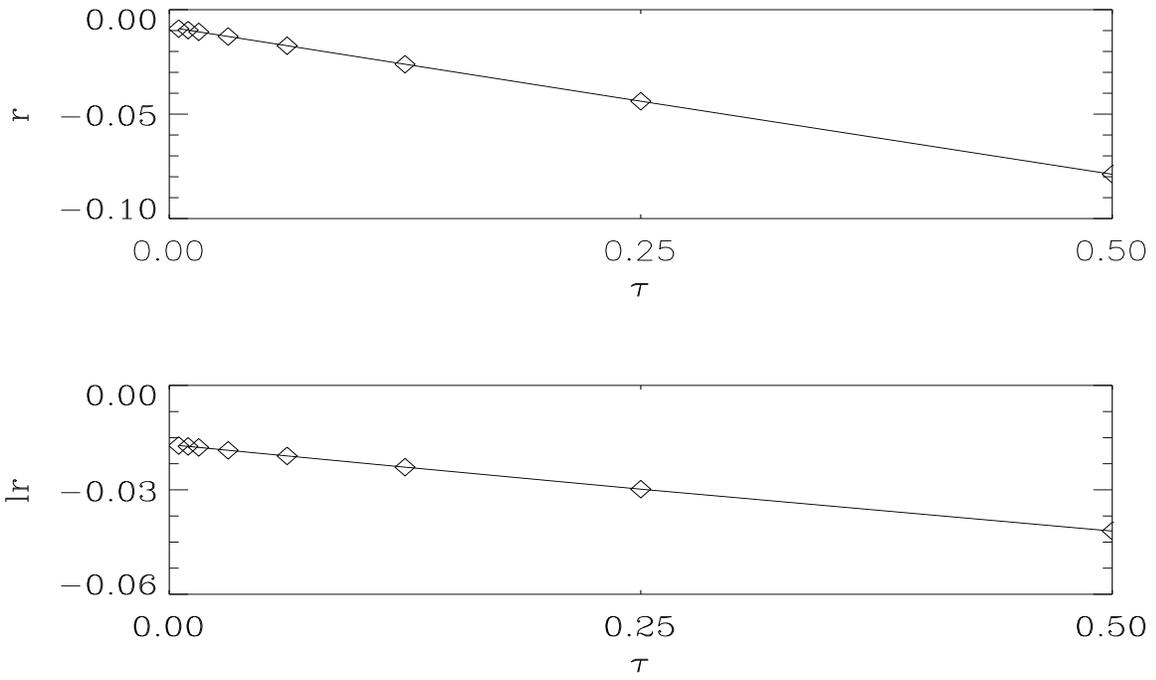

Figure 3: The parameters $r$ and $l_r$ defined in the text as identified with various values of the time step $\tau$. The smallest integration time is $2\ 10^{-4}s$ which gives 2400 integration points by oscillation period.

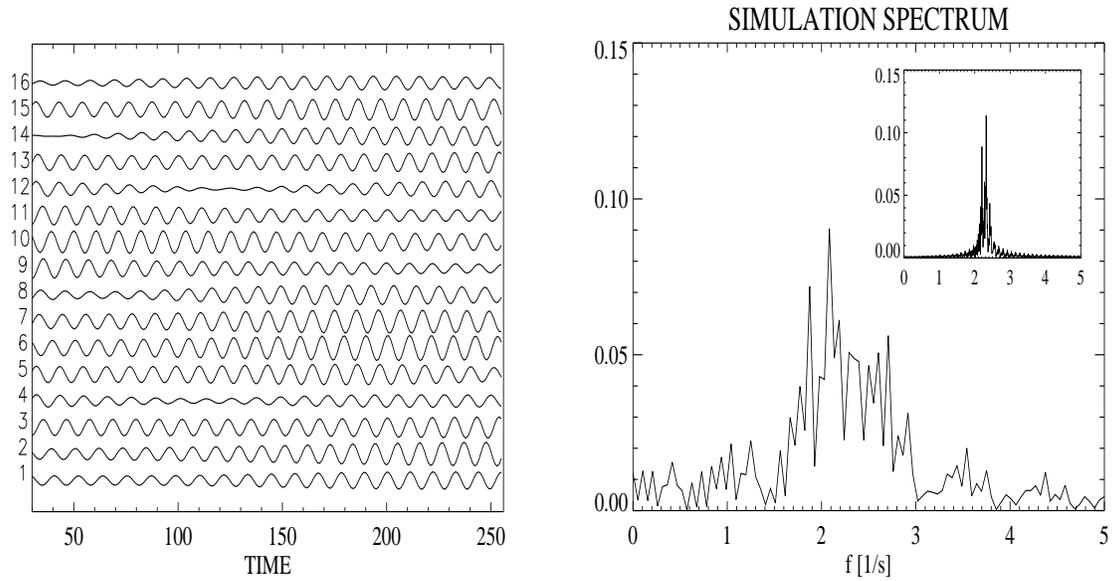

Figure 4: (a) Simulation of the continuous model. The parameters used correspond to the smallest $\tau$ in Figure 3. Initial conditions are provided by the experimental data. (b) Power spectrum of the simulated signal computed over the identification window. The inset shows the spectrum computed over the entire time series.

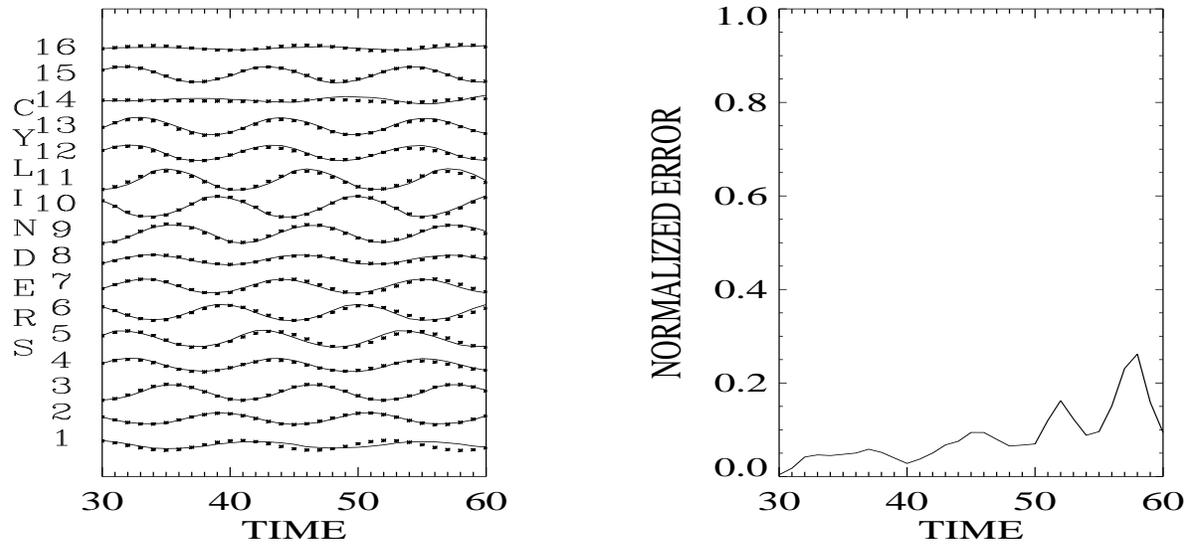

Figure 5: (a)Over plot of the coupled Landau equation simulation (dotted line) and experimental data (solid line)   (b) the normalized error as defined in the text.

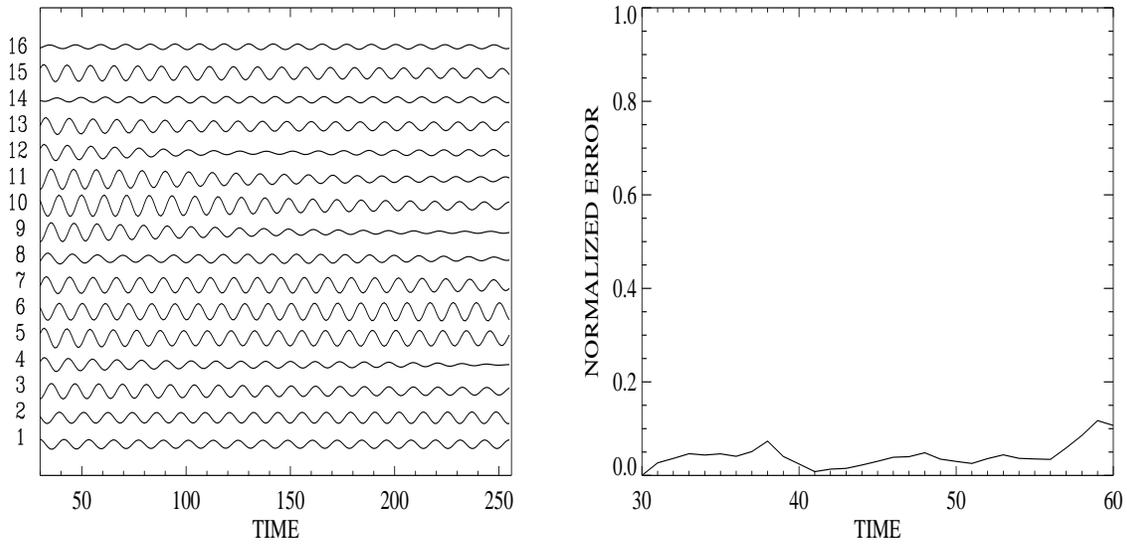

Figure 6: (a)Simulation over the full range of the data set using the parameters of the 22 parameters optimization. (b)Short time normalized error in the training window.

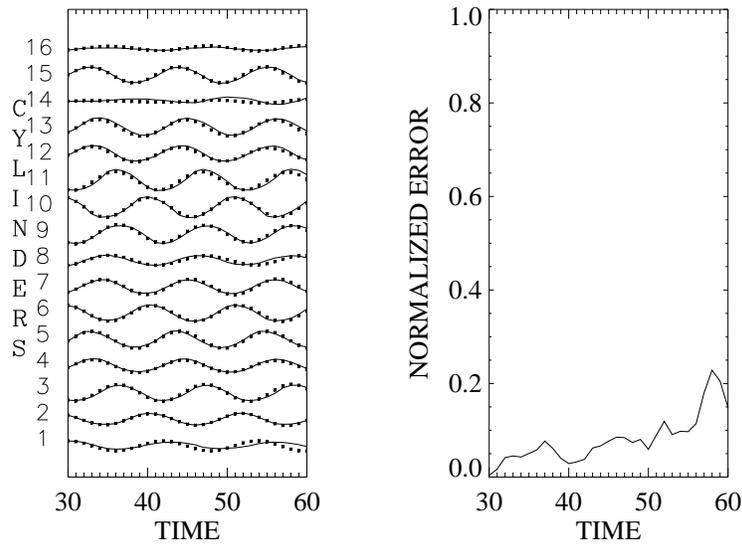

Figure 7: (a) Over plot of the simulation with the extra term $|A|^4 A$ added to the coupled Landau equation in the first training window. The dotted line are simulations and solid lines are experimental data. The parameters averaged from training windows from 30 to 90 are used. (b) Short time normalized error in the training window.

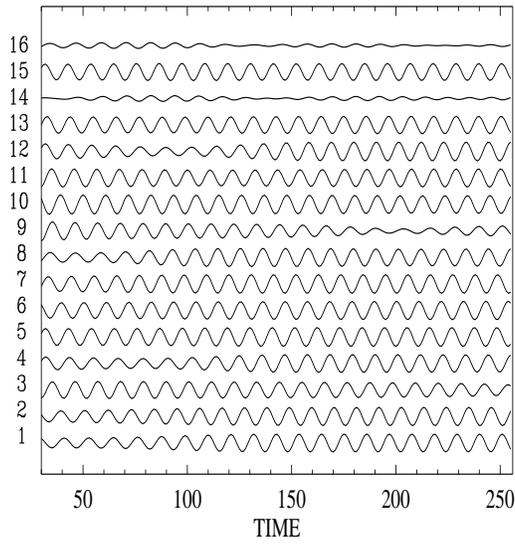

Figure 8: Simulation over the entire duration of the data set with the extra term $|A|^4 A$.

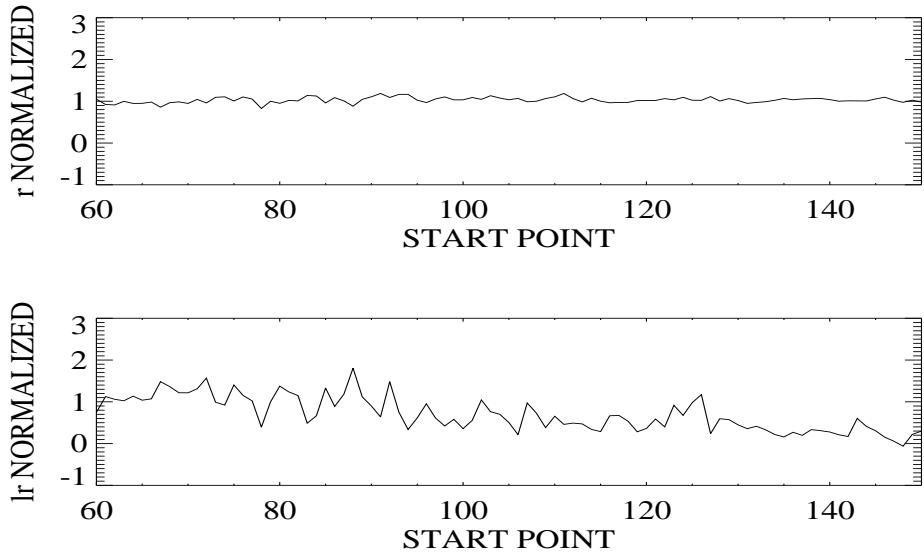

Figure 9: Parameters $r$ and $l_r$ as function of the starting point of the optimization window. The identification is performed in a window of 30 time steps taken from experimental data set.

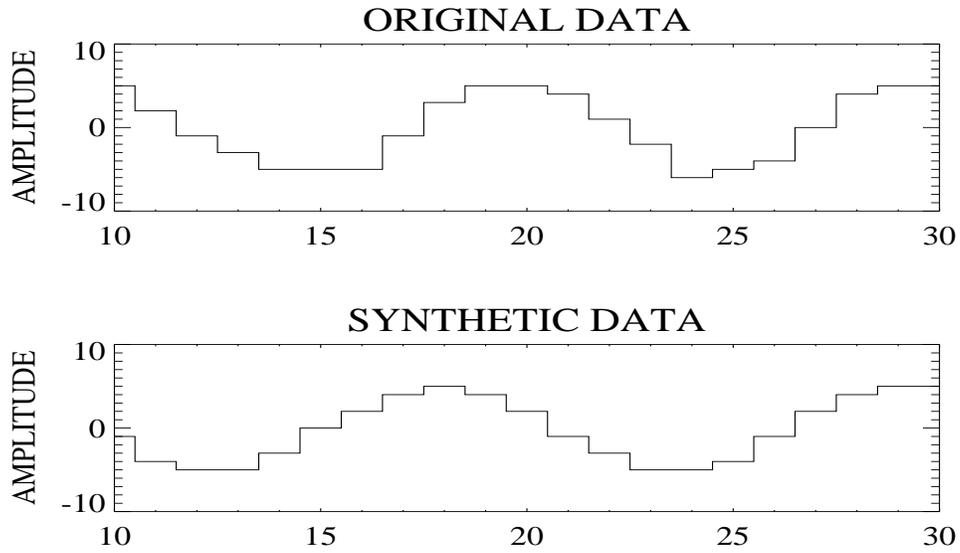

Figure 10: (a) Experimental data showing the discreteness of the signal due to the measurement process. (b) Synthetic data where the discretisation or "round-off" parameter S is set to 6.

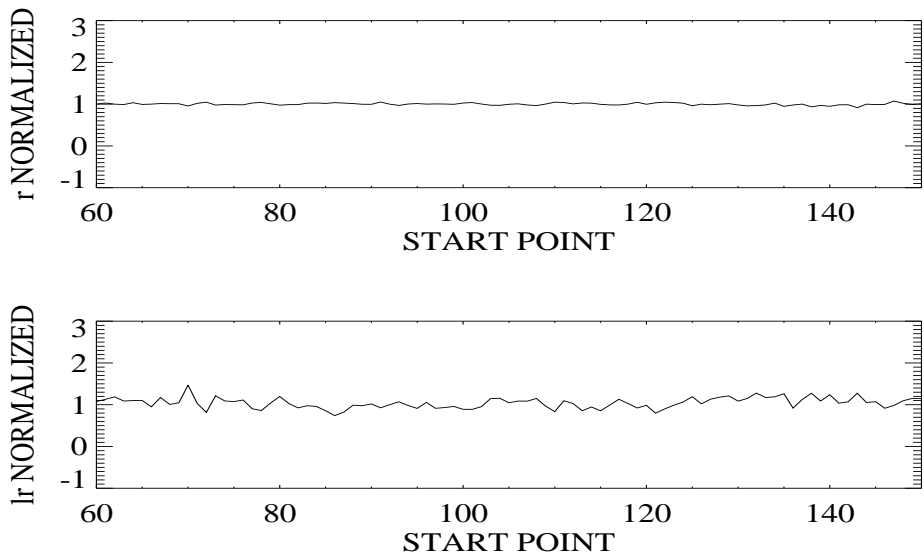

Figure 11: Normalized parameters $r_{NORMALIZED} = \hat{r}/r$ and $l_{rNORMALIZED} = \hat{l}_r/l_r$ as functions of the starting point of the optimization window when the round-off parameter $S = 6$

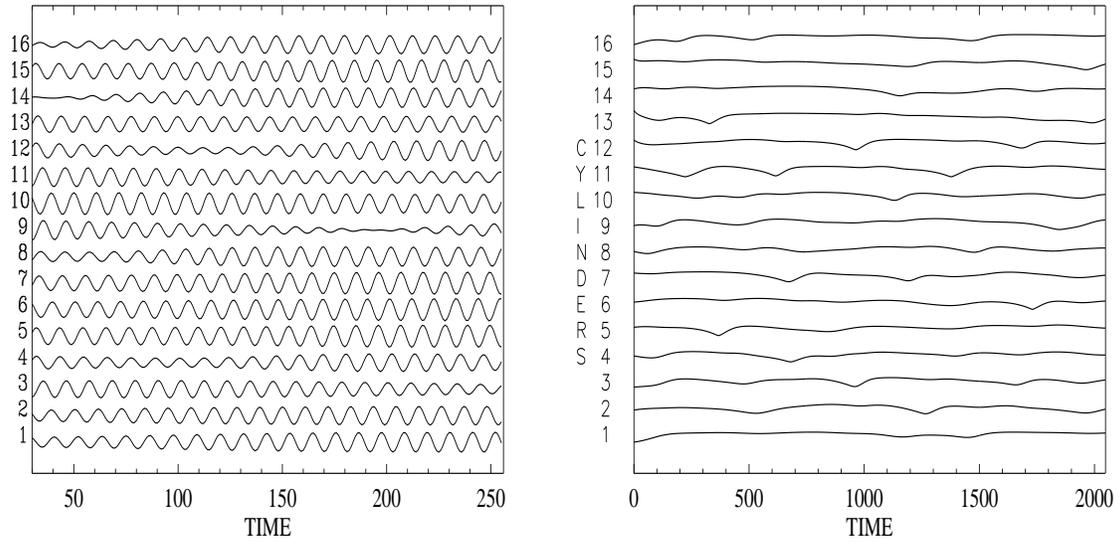

Figure 12: (a) Simulation of the large-time-step model where the integration timestep coincides with the measurements ($\tau = T$) and (b) The moduli $|A_i|$ in a simulation pursued over times much longer than the duration of the experimental series. Note that the intermittency, characterized by holes in the moduli persists over the entire simulation.

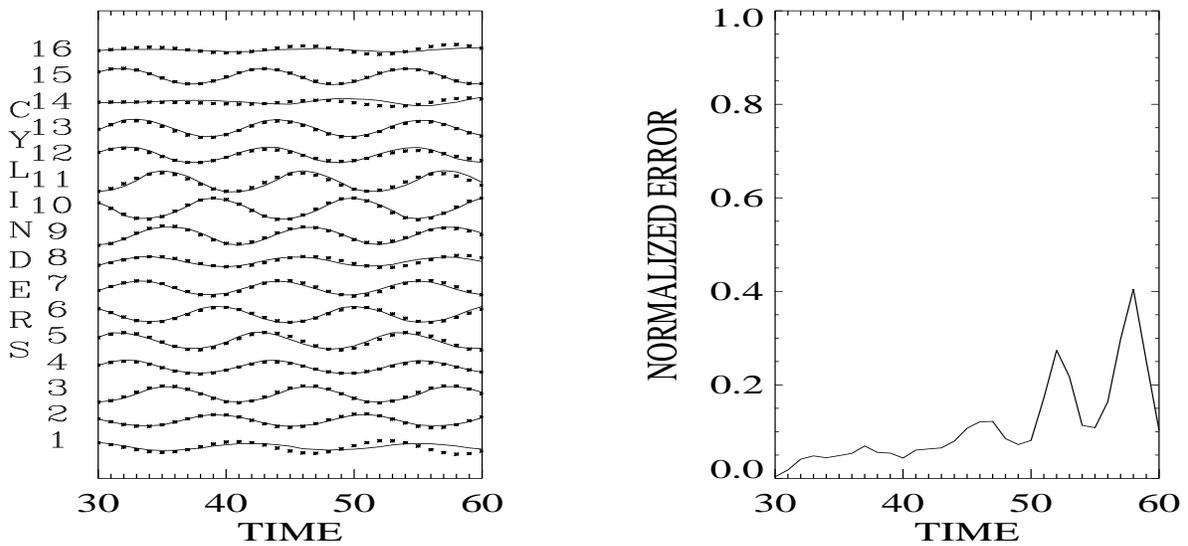

Figure 13: (a) Over plot simulation (dotted line) for the large-time-step model and original data (solid line) (b) Normalized error for the optimization window.

25